# Transfer of Energy and Momentum between Magnetoactive Surface Microstructure and a Solid Object

*Arne Geldof, Jan Kopačin, Izidor Straus, Raphael Kriegl, Gaia Kravanja, Luka Hribar, Matija Jezeršek, Mikhail Shamonin, Gašper Kokot, and Irena Drevenšek-Olenik*[*]

A. Geldof, J. Kopačin, I. Straus, G. Kokot, I. Drevenšek-Olenik

University of Ljubljana, Faculty of Mathematics and Physics, Jadranska 19, SI-1000, Ljubljana, Slovenia

R. Kriegl, M. Shamonin

East Bavarian Centre for Intelligent Materials (EBACIM), Ostbayerische Technische Hochschule (OTH) Regensburg, Seybothstr. 2, 93053 Regensburg, Germany

G. Kravanja, L. Hribar, M. Jezeršek

University of Ljubljana, Faculty of Mechanical Engineering, Aškerčeva 6, SI-1000, Ljubljana, Slovenia

G. Kokot, I. Drevenšek-Olenik

Department of Complex Matter, J. Stefan Institute, Jamova 39, SI-1000, Ljubljana, Slovenia
E-mail: irena.drevensek@ijs.si

**Funding**: Slovenian Research and Innovation Agency (ARIS) under research programs P1-0192 and P2-0231, European Union's Horizon Europe research and innovation programme under Marie-Skłodowska-Curie Actions Doctoral Network MAESTRI (grant agreement No. 101119614), Slovenian Research and Innovation Agency (ARIS) and German Academic Exchange Service (DAAD) under Slovenian-German bilateral project BI-DE/25-27-003, i.e., project No. 57753025, German Research Foundation (DFG) under project No. 524921900, and the GREENTECH project, co-financed by the European Union – NextGenerationEU.

**We investigate the physical mechanisms driving directional transport of solid objects by micro-lamellar structures laser-inscribed on the surface of a magnetoactive elastomer (MAE). When subjected to a rotating magnetic field with magnitude of 175 mT and a time period of 0.4 s, the lamellas reorient within a few milliseconds, reaching angular velocities up to 1100 rad s$^{-1}$. This rapid motion is crucial for efficient momentum and energy transfer to objects in contact with the lamellas. The analysis of collisions of a single lamella with a lead ball with a 2.2 mm diameter shows that the lamella can transfer around 50 nJ of energy, propelling the ball to a speed of around 35 mm s$^{-1}$. We show how this value sets the upper limit for the ball's transport speed on multi-lamellar MAE arrays. We also explain the background of three distinct transport regimes (kicking, pushing, and bouncing modes) observed on these magnetically driven "conveyor belts".**

## 1. Introduction

Actuating systems based on magnetic materials offer significant advantages in applications where contactless operation and minimal interference with biological processes are essential. Among various possibilities, composites consisting of an elastomeric matrix and a magnetic filler - commonly referred to as magnetoactive elastomers (MAEs), magnetorheological elastomers, or simply magnetic elastomers - have recently gained considerable attention due to their versatility and strong responsivity to a magnetic field.[1-4] It has been demonstrated that surface microstructures made from MAEs, most often designed to resemble biological cilia (magnetic artificial cilia (MAC)), can effectively manipulate liquid flow in microfluidic systems.[5-8] They have also proven useful for transporting gas bubbles in liquids and liquid droplets in gaseous environments,[9] a key feature in magnetic digital microfluidics.[10]

While the transportation and manipulation of bubbles and droplets via MAC assemblies have been extensively studied, their application for transporting solid objects is still far from being systematically explored. This fact stems from the wide range of possible system parameters, including variations in the shape, size, and actuation modes of magnetic microstructures and differences in the shape, size, composition, and surface interactions of the solid objects used across different studies. As a result, resolving similarities and discrepancies between reported findings remains challenging.

Cilia-inspired two-dimensional (2D) arrays of periodically arranged hair-like actuating structures (e.g., pillars, cylinders, cones) exposed to a time-varying magnetic field have been shown to enable controlled transport of objects with varying shapes and sizes.[11-17] The symmetry breaking required for directional transport was achieved either through asymmetric structural elements or via asymmetric actuation methods. Additionally, multidirectional particle transport was demonstrated by exploiting a combination of rotational and translational motion of the actuating magnets.[18,19]

Although less common in nature, unidirectional object transport can also be achieved using one-dimensional (1D) arrays of planar microstructures. These structures are referred to in the literature by various terms, such as micro wall-, micro plate-, and micro lamellar-arrays. Due to their relatively larger mass, individual actuation elements of 1D arrays facilitate greater momentum transfer to transported particles, enabling higher transport velocities.[20] Reported maximum transport speeds at present reach approximately 100 mm s$^{-1}$.[21] Such magnetic "conveyor belts" can generate multi-substance transport in both gas and liquid environments, and are therefore interesting for versatile applications.[22]

Our recent work reported the efficient transport and separation of solid objects on a magnetoactive micro lamellar structure fabricated via a laser micromachining method.[23] Object propulsion was achieved by exposing the assembly to a rotating quadrupole magnet. Different transportation modes were observed at different rotational frequencies. Building upon these findings, this work aims to elucidate the underlying mechanisms of directional object transport on magnetic micro lamellar structures by analyzing momentum and energy transfer processes between the microstructure and a spherical microparticle. By examining the dynamic response of lamellas to a time-varying magnetic field and investigating in detail the collision process between a single lamella and a spherical particle, we uncover the fundamental principles governing the transport of solid objects by 1D magnetically actuated surface systems. To the best of our knowledge, thus far, no quantitative analysis of the transferred energy and momentum from an MAE-based magnetoactive structure has been reported.

## 2. Results

### 2.1. Response of Lamellas to Rotating Magnetic Field

**Figure 1**a presents a schematic drawing of the investigated multi-lamellar MAE structure and the associated coordinate system. The height ($H$) and width ($W$) of the MAE lamellas, in both single-lamellar and multi-lamellar structures, were approximately 400 μm and 70 μm, respectively. The periodicity ($P$) of the multi-lamellar assembly was approximately 390 μm. The lamella length ($L$) was 20 mm. The total length of the multi-lamellar assembly along the x-axis was also approximately 20 mm, typically comprising 50 lamellas per sample. The polyethylene terephthalate (PET) substrate beneath the lamellar structure had a thickness of around 200 μm. The sample was fixed to a glass support and mounted above the quadrupole magnet.

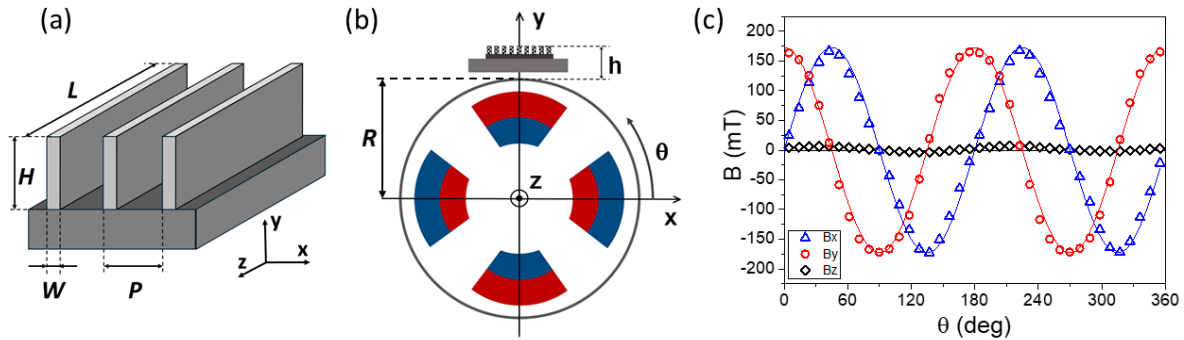

**Figure 1.** (a) A schematic drawing of the lamellar structure, showing the characteristic parameters and the designated coordinate system. (b) A schematic illustration of the lamellar structure assembled above the quadrupole magnet assembly incorporated inside the cylindrical frame with diameter 2$R$. (c) The magnetic flux density components, $B_x$, $B_y$, and $B_z$, at the sample position as a function of rotation angle $\theta$ of the quadrupole magnet. The data shown corresponds to a height of the sample $h = 2$ mm above the magnet.

The experiments started with measurements of the magnetic field at the sample position. Figure 1b illustrates a placement of the sample above the magnet assembly. The latter was enclosed within the cylindrical frame with a diameter 2$R$ of 42 mm and a length of 72 mm (along the $z$-axis). The sample was positioned at the midpoint of the cylinder's length. The magnetic field strength was varied by varying the distance $h$ of the sample from the cylinder surface. Figure 1c depicts measured dependencies of the three magnetic flux density B components on the rotation angle $\theta$ of the cylinder for $h = 2$ mm. The orientation $\theta = 0$, where the magnetic field is oriented parallel to the y-axis, corresponds to Figure 1b. The symbols represent the measured values, while solid lines represent fits to the harmonic functions:

$$B_x = B_0 \sin 2\theta, \quad B_y = B_0 \cos 2\theta \quad , \tag{1}$$

with $B_0 = 175 \pm 8$ mT, signifying a magnetic field with a constant magnitude, whose direction rotates (clockwise) at a constant rate within the $xy$-plane. The rotation rate d$\theta$/d$t$ is determined by the rotation period $t_{0q}$ of the magnet assembly. The corresponding frequencies (1/$t_{0q}$) used in the experiments ranged from 0.1 to 10 Hz.

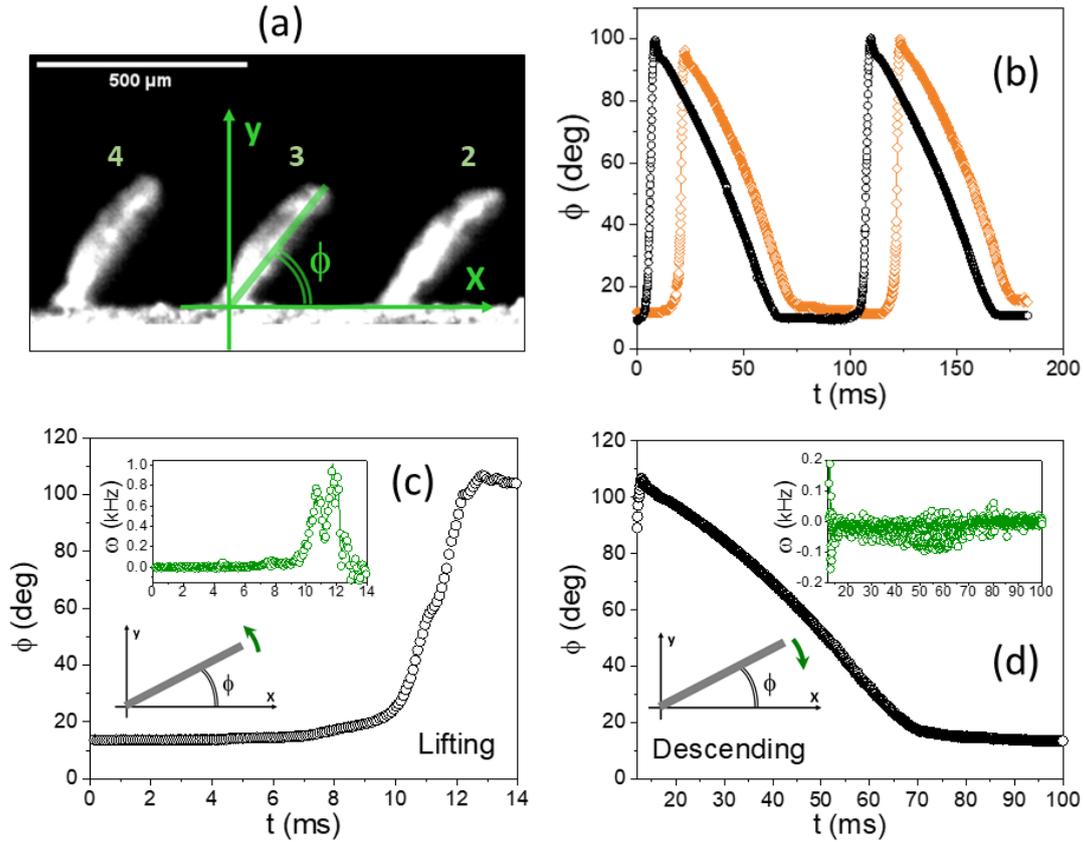

**Figure 2.** (a) Side-view image showing three lamellas in the central region of the sample, with the coordinate system and the effective inclination angle $\Phi$ indicated. The scale bar is 500 μm. (b) Time dependence of inclination angle $\Phi$ for two lamellas during rotation of the quadrupole magnet, which had a rotation period $t_{0q}$=400 ms. The time delay between lifting of different lamellas (the shift between black and orange data points) is $N \cdot P/v_f$, where N=1,2,3…(the data are shown for N=6) and $v_f$ is the fringe velocity determined by the rotation frequency of the magnet. (c) Time dependencies $\Phi(t)$ and $\omega(t)=d\Phi/dt$ (inset) during the lifting phase of the middle lamella. (d) Time dependencies $\Phi(t)$ and $\omega(t)=d\Phi/dt$ (inset) during the descending phase of the middle lamella. The transition from the lifting to the descending phase occurs approximately at $t$ = 13 ms.

The initial investigation focused on the dynamic response of MAE lamellas to the rotating magnetic field. For this purpose, a 3 mm wide central region of the sample was monitored with a high-speed camera from the side view, i.e., along the *z*-axis (Figure 1b and **Figure 2**a). Correspondingly, the magnetic field-induced bending of seven lamellas was video-monitored (Video S1 in [24]). During rotation of the quadrupole magnet, the lamellas exhibited a distinctive punching-like motion: a rapid lift from a nearly horizontal to a vertical orientation, followed by a much slower descent back to the horizontal position. This motion, resembling the characteristic "power stroke" and the "recovery stroke" of the cilia, repeated four times per rotation period of the magnet, indicating that the response was independent of the magnetic field's polarity.

The video analysis results of the lamellas' motion are shown in Figures 2b-d. Figure 2b presents the time dependence of the effective inclination angle $\Phi$ for the first and seventh lamella in the row, tracked over approximately two periods of their movement. Figures 2c-d provide a more detailed view of the lifting and descending phases of the central lamella (lamella no. 4), along with the corresponding angular velocities, $\omega = d\Phi/dt$. The rotation period $t_{0q}$ of the quadrupole magnet was 400 ms, while the response period of the lamellas $t_{0l}$ was 100 ms. Analysis of 14 reorientation events across seven lamellas yielded an average lifting time $\tau_L = 2.7 \pm 0.3$ ms, defined as the time required for $\Phi$ to increase from 10% to 90% of its total change. The average descending time was $\tau_D = 43.5 \pm 1.9$ ms, representing the time for $\Phi$ to decrease from 90% to 10% of its total change. The corresponding average angular velocities during the lifting and descending phases are $<\omega_L> = 464 \pm 53$ s$^{-1}$ and $<\omega_D> = -28.8 \pm 1.2$ s$^{-1}$ (where s$^{-1}$ means rad s$^{-1}$), respectively, while the maximal observed values of $\omega$ during 14 lifting events are $\omega_{max} = (d\Phi/dt)_{max} = 1110 \pm 200$ s$^{-1}$. A large difference between $\tau_L$ and $\tau_D$ results in the generation of a step-like surface formation ("surface fringe"), separating the regions with more or less vertically oriented lamellas from the areas with horizontally oriented lamellas.

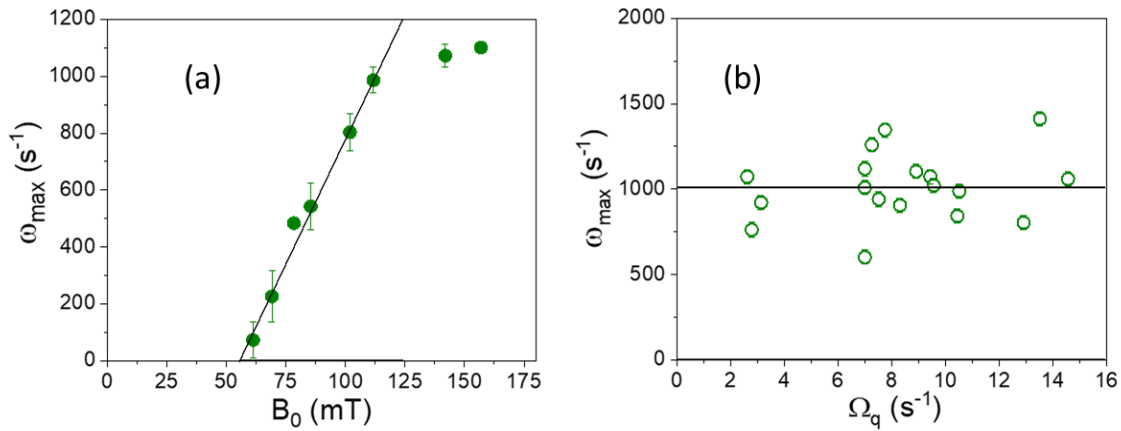

**Figure 3.** (a) Dependence of $\omega_{max} = (d\Phi/dt)_{max}$ observed during the lifting phase of the lamellas on magnetic field strength. The line corresponds to a linear fit for the fields below 125 mT. (b) Dependence of $\omega_{max}$ on rotational velocity $\Omega_q = 2\pi/t_{0q}$ of the quadrupole magnet. The line corresponds to the average value of 1010 s$^{-1}$.

Next, we investigated how the lifting dynamic of the lamellas depends on the magnetic field strength $B_0$. To do this, we varied the height $h$ of the sample above the quadrupole magnet. **Figure 3**a shows the observed maximum angular velocity $\omega_{max}$ as a function of $B_0$. Magnetic fields below 60 mT did not induce any bending of lamellas. Within the range 60 mT $< B_0 <$ 110 mT, $\omega_{max}$ detected during the lifting process increases approximately linearly with increasing field strength. For $B_0 >$ 110 mT, a tendency of $\omega_{max}$ toward saturation emerges. We also investigated how the magnet's rotation speed influences the lifting velocity at $B_0 = 175$ mT. Measurements were performed in several sample regions. The results are collected in Figure 3b. They indicate that the values of $\omega_{max}$, which are about two orders of magnitude higher than the angular velocity of the rotating quadrupole magnet $\Omega_q = 2\pi/t_{0q}$, are independent of $\Omega_q$.

## 2.2. Rebound of a Rigid Ball from a Single Lamella

To investigate the interaction process between the solid balls and a single lamella, we sculpted out from an MAE film a specific sample with only one lamella with the dimensions $H = 400$ µm, $W = 70$ µm, and $L = 20$ mm. During the experiments, the substrate was tilted at an angle $\alpha$ between 2° and 8° relative to the horizontal plane (**Figure 4**a). This tilt allowed the balls to acquire an initial velocity as they rolled along the substrate and subsequently collided with the lamella in its punching phase. We investigated collisions at initial ball velocities ranging from 5 to 30 mm s$^{-1}$.

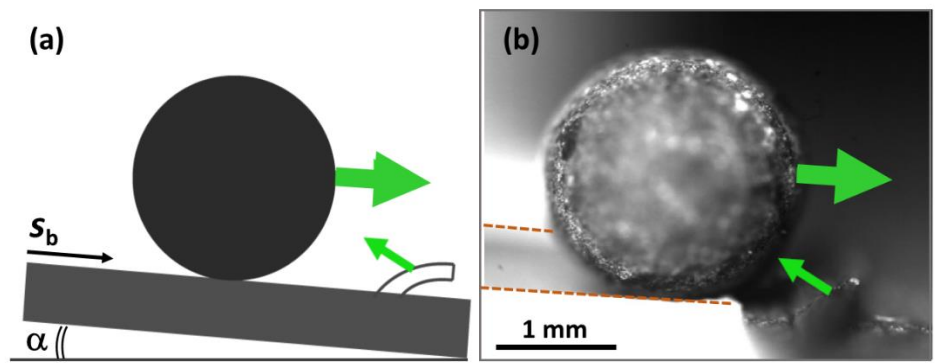

**Figure 4.** (a) Schematic drawing of the ball rolling down the inclined substrate toward the MAE lamella in its "power stroke" phase. (b) Photograph of the lead ball approaching the lamella. The dashed lines represent the guiding channel used to direct the ball towards the lamella. The thick arrow indicates the direction of the ball's velocity, while the thin arrow represents the bending direction of the lamella.

A surface channel made of paper was utilized to constrain the ball's trajectory and ensure its impact along the lamella's bending direction (see dotted lines in Figure 4b). We tested balls of various materials (wood, lead, plastic, and concrete) with diameters between 1 and 8 mm and masses between 0.03 and 1 g. Lead balls were selected for detailed analysis, as their surface irregularities made their motion, especially rotation, easier to track using motion analysis software. The final choice of ball size (diameter of approximately 2.2 mm, mass of about 0.07 g, Figure 4b) was made to ensure optimal conditions for analysis of the collision dynamics (Video S2 in [24]). Larger balls tended to roll over the lamella without producing a proper collision, while smaller balls bounced off too quickly, making precise measurements quite difficult.

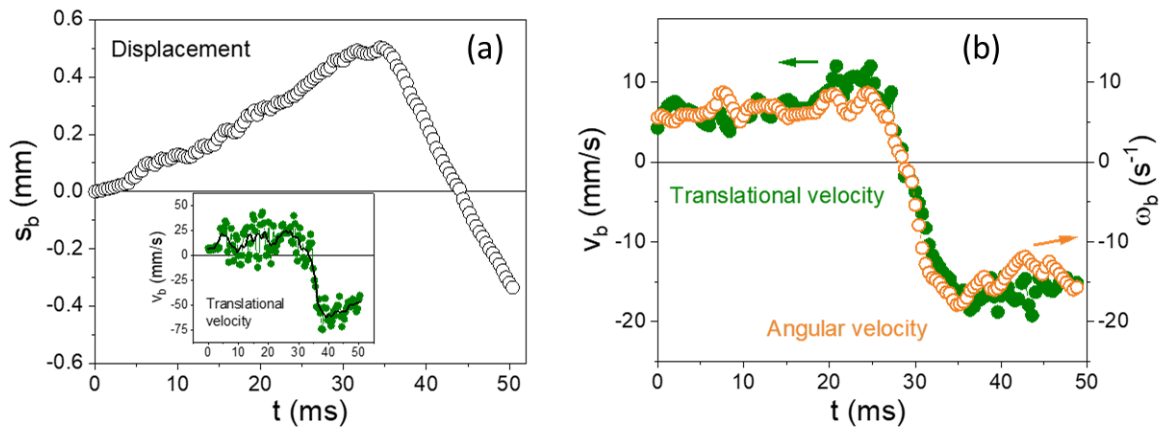

**Figure 5.** (a) Time dependence of the ball's displacement sb (see Figure 4a) during the collision event with the punching lamella. The initial velocity of the ball is about 20 mm s$^{-1}$. The collision starts about 30 ms after the start of observation. The inset shows translational velocity, $v_b$=d$s_b$/d$t$, of the ball as a function of time. The black solid line represents the averaged data. (b) The collision dynamics for the ball with an initial velocity of about 5 mm s$^{-1}$. Only the averaged data are shown. Green solid circles denote translational velocity, and yellow open circles give the angular velocity of the ball. Their ratio reveals that the ball is rolling on the surface without slipping.

**Figure 5** shows the results of two collisions between the ball and the lamella during its punching phase induced by a rotating magnetic field with $B_0$ = 175 mT. In the first collision, the ball had an initial velocity of approximately 20 mm s$^{-1}$. This collision began around 30 ms after the start of the observation. The inset illustrates the translational velocity of the ball, $v_b$=d$s_b$/d$t$, as a function of time. It reveals that the ball was rebounded from the lamella with a velocity roughly two times larger than its initial velocity. In the second collision, the ball's initial velocity was about 5 mm s$^{-1}$. For this case, the changes in the ball's angular velocity, $\omega_b$ = d$\varphi_b$/d$t$, where $\varphi_b$ is the rotation of the ball around its center of mass, were tracked and are shown in Figure 5b. It can be observed that the ratio between $v_b$ and $\omega_b$ is slightly over 1 mm, corresponding to the diameter of the ball. This confirms that the ball was rolling without slipping on the supporting surface both before and after the collision. The relatively high static friction between the ball and the surface needed for its rolling is attributed to a very thin MAE layer that remained adhered to the PET substrate following the laser micromachining process.

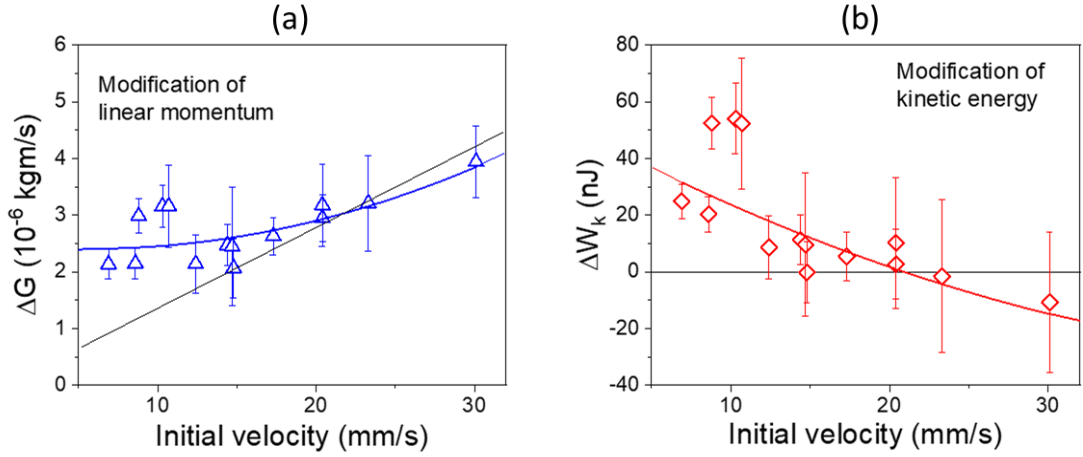

**Figure 6.** (a) Magnitude of collision-induced modification of linear momentum of the ball $\Delta G$ as a function of its initial velocity ($v_{bi}$). (b) Collision-induced modification of kinetic energy of the ball $\Delta W_k$ as a function of its initial velocity. The thin (black) lines correspond to the case of totally elastic reflection of the ball from a solid wall. The thick lines are fits to the 2$^{nd}$ order polynomial.

The results from the analysis of 14 collision events are summarized in **Figure 6**. In every investigated case, after colliding with the punching lamella, the ball reversed direction relative to its initial motion. Video analysis revealed that the physical contact between the lamella and the ball lasted for $\Delta t_c = 12.0 \pm 2.7$ ms, practically independent of the initial velocity. Figure 6a shows the magnitude of modification of linear momentum of the ball $\Delta G = m_b(v_{bf}+v_{bi})$, where $m_b$ is the mass of the ball and $v_{bf}$ and $v_{bi}$ are the magnitudes of the final and initial velocities of the ball, respectively, as a function of the initial velocity of the ball $v_{bi}$. These velocity values were averaged over short time intervals - just a few milliseconds - before and after the collision. The thin (black) line represents the expected momentum change for a conventional elastic bouncing from a rigid wall, which would yield $\Delta G = 2m_b v_{bi}$. Notably, for initial velocities below 20 mm s$^{-1}$, the ball gains additional momentum. By fitting the data to the 2$^{nd}$ order polynomial, this excess can be extrapolated to a value of $\Delta G_0 = (2.5 \pm 0.7) \cdot 10^{-6}$ kg m s$^{-1}$ for $v_{bi} = 0$, which results in the generated speed of the ball $v_{bg} = v_{bf}(v_{bi}=0) = \Delta G_0/m_b = (36 \pm 10)$ mm s$^{-1}$ induced by the punching of the single lamella. Accordingly, the average force exerted by the lamella on a stationary ball during the collision is estimated as $<F> = \Delta G_0/\Delta t_c = (3.0 \pm 1.5) \cdot 10^{-4}$ N.

Figure 6b presents the change in the ball's total kinetic energy, defined as $\Delta W_k = (7/10)m_b(v_{bf}^2 - v_{bi}^2)$, plotted as a function of the initial velocity of the ball $v_{bi}$. The data indicate that for initial velocities below approximately 20 mm s$^{-1}$, the ball gains kinetic energy, which extrapolates to $\Delta W_{kg} = \Delta W_k(v_{bi}=0) = (51 \pm 24)$ nJ (see thick red line in Figure 6b) - consistent with the momentum increase observed in Figure 6a. Taking into consideration the above calculated average force ⟨$F$⟩ that the lamella exerts on the ball, one obtains the interaction distance $s = \Delta W_{kg}/⟨F⟩ = (170 \pm 160)$ μm, which aligns well with the expected value $s \leq H = 400$ μm.

For comparison, we also analyzed collisions between the ball and the lamella held in a fixed vertical position ($\Phi = 90°$, see Figure 1) stabilized using a stationary magnetic field of 175 mT applied along the y-axis, i.e., perpendicular to the substrate. In this configuration, the magnitude of final velocity $v_{bf}$ was typically around 50 % of the magnitude of initial velocity $v_{bi}$, indicating that the collision with the stationary lamella is highly inelastic. This result further highlights the significant role of the lamella's punching motion in altering the ball's dynamics.

## 2.3. Transport of a Ball by a Multilamellar Array

The final part of our investigation focused on the transport properties of the lead ball after being placed on the top of a multilamellar array. Consistent with our previous study involving glass balls,[23] we observed that the transport properties strongly depend on the velocity $v_f$ of a surface fringe governed by the angular velocity of the rotating quadrupole magnet as $v_f=(R+h)\Omega_q$. Depending on the value of $v_f$, two different transport regimes were analysed: the so-called "pushing mode" at lower $v_f$ values and the "bouncing mode" at higher $v_f$ values.

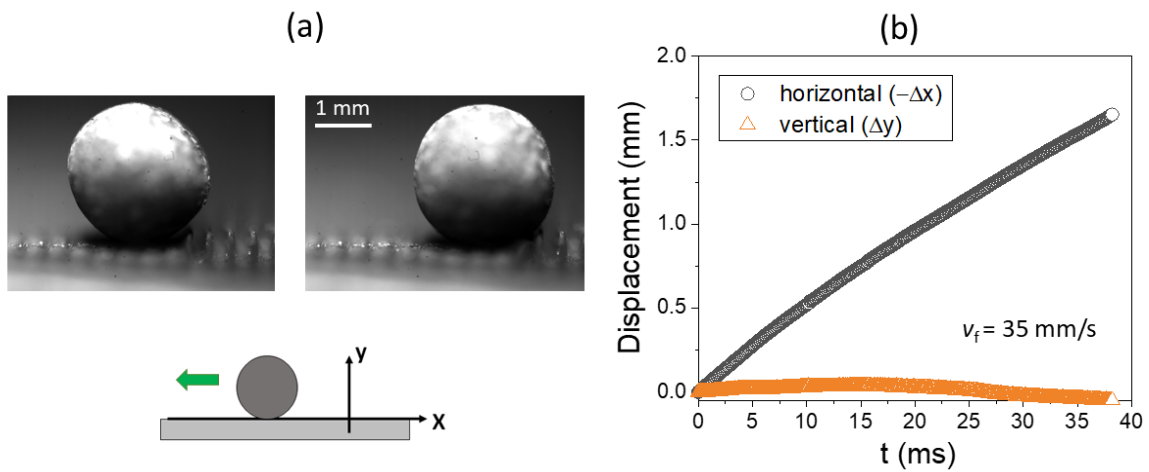

**Figure 7.** (a) Sequence of images captured during the ball's transportation on the multilamellar structure in the "pushing mode". In the time lag between the images, the ball moved to the left for around one distance between the neighboring lamellas ($P = 0.39$ mm). (b) Time dependencies of horizontal ($-\Delta x$) and vertical ($\Delta y$) displacements of the center of the ball during its 1 mm travel along the surface. The deviations from $\Delta y = 0$ are attributed to irregularity of the surface structure and deviations of the ball's shape from a perfect sphere.

**Figure 7** shows an example of the transport in the "pushing mode" (Video S3 in [24]). In this regime, the ball is effectively pushed forward by the moving surface fringe, resulting in an average horizontal transport speed $\langle v_b \rangle = |\Delta x/\Delta t| = 43 \pm 9$ mm s$^{-1}$, which is, within the experimental error, the same as the fringe velocity $v_f = 35 \pm 6$ mm s$^{-1}$.

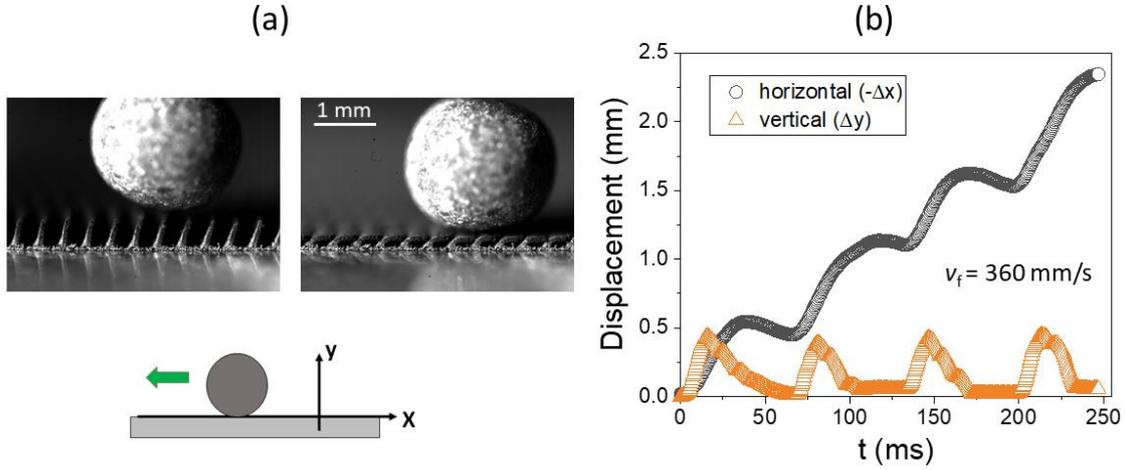

**Figure 8.** (a) Sequence of images captured during the ball's transportation on the multilamellar structure in the "bouncing mode". In the time lag between the images, the ball moved to the left for around one distance between the neighboring lamellas ($P = 0.39$ mm). (b) Time dependencies of horizontal (-$\Delta x$) and vertical ($\Delta y$) displacements of the center of the ball during its 2.5 mm travel along the surface. The deviations from $\Delta y = 0$ are attributed to vertical jumps of the ball induced by an overtaking "surface fringe".

**Figure 8** illustrates an example of the transport in the "bouncing mode" (Video S4 in [24]). In this regime, the fringe periodically overtakes the ball, causing it to execute upward jumps over the fringe step each time a new fringe arrives. This repeated process results in an average horizontal transport speed $\langle v_b \rangle = |\Delta x/\Delta t| = 9.5 \pm 0.4$ mm s$^{-1}$, which is considerably lower than the fringe speed ($v_f = 360 \pm 11$ mm s$^{-1}$) and also much lower than the speed obtained in the "pushing mode".

The transition between the two regimes was observed at $v_f \sim 50$ mm s$^{-1}$, which is about 40% higher than the calculated generated speed of the ball, $v_{bg} \sim 36$ mm s$^{-1}$, induced by the impact of the single punching lamella on the resting ball. This discrepancy is attributed to the geometric relationship between the lamellas and the ball: the lamellar spacing ($P = 0.39$ mm, see Figure 1) is smaller than the ball radius ($R_b = 1.1$ mm). Consequently, in the pushing regime, more than one lamella can simultaneously interact with the ball during the collisions, leading to a higher effective propulsion force and, therefore, an increased generated speed.

## 3. Discussion and Conclusions

Our results show that by resolving the momentum and energy exchange between a single magnetoactive lamella and the object to be transported, one can predict this object's maximum achievable transportation speed on a multilamellar system. This speed corresponds to the ideal "pushing mode" scenario, in which the object (e.g., a ball), upon collision, gains the same velocity as the velocity of the lamellar fringe. The details of the collision process depend on magnetic-field-induced deformation (bending) of the lamellas. Larger speeds can therefore be achieved by storing larger elastic deformational energy. However, this approach evidently involves a trade-off: higher elastic energy requires stronger magnetic fields to induce sufficient bending. Consequently, considering specific limitations associated with the magnetic field, optimizing the size and shape of the lamellas becomes crucial and must be adapted to the mass and dimensions of the transported object. A detailed computational analysis of this optimization problem is beyond the scope of the present work and will be presented in a separate publication.

When the speed of the lamellar fringe $v_f$ exceeds the velocity imparted to the object during collisions, the fringes begin to "run" beneath the object, effectively overtaking it. This results in vertical jumping of the object and intermittent loss of contact with the transporting surface, which reduces the efficiency of transport and decreases overall transport speed in the horizontal ($x$) direction. In our system, governed by the selected size of the ball and the pitch of the lamellar structure, the number of collisions per jump was observed to be up to five. Some of these collisions produce forces that counteract gravity, effectively redirecting the ball upward. For the example shown in Figure 8, where single jumps correspond to vertical displacement of $\langle \Delta y \rangle = 0.39 \pm 0.03$ mm, the corresponding potential energy gain is $\Delta W_p = 268 \pm 21$ nJ. Within the experimental error, this value agrees with the expected energy gain from five collisions, $5\Delta W_{kg} = 255 \pm 120$ nJ, suggesting that momentum transfer in the "bouncing" mode occurs predominantly in the vertical ($y$) direction. This explains why this mode of transportation is less efficient.

In addition to the above-described "pushing" and "bouncing" modes, as reported in our previous work, there also exists a third transport mode in which the object, after colliding with the lamellar fringe, momentarily moves faster than the fringe itself.[23] We propose to call this mode the "kicking" mode. In this mode, the object bypasses the next lamella without a collision and continues moving forward until surface friction gradually dissipates its energy, eventually bringing it to a stop. The object then "waits" at rest until the fringe catches up, resulting in an average transport velocity of $\langle v_b \rangle = v_f$. As the fringe velocity $v_f$ increases, collisions may begin to occur before the object comes to a complete stop. At some value of $v_\text{f}$, the energy injected during collisions balances the energy lost due to friction, leading to a transition into the "pushing" mode.

In summary, the demonstrated discrete "injections" of energy and momentum are a key mechanism of object transport on microstructured MAE surfaces. When combined with surface friction (for solid objects), capillary adhesion (for liquid droplets in a gas or bubbles in a liquid), and the influence of gravity (e.g., on inclined surfaces), this mechanism provides a robust foundation for designing a wide range of magnetic-field-controlled micro-transport platforms.

## 4. Experimental Section

*Fabrication of MAE films on PET substrate*

A base PDMS polymer mixture was created by combining polymers VS 100000 and MV 2000, Modifier 715, all obtained from Evonik Industries AG, and a silicone oil AK10 from Wacker Chemie AG. To obtain the MAE, this base polymer was combined with carbonyl iron powder (CIP SQ from BASF) and stirred until homogeneous. Then, a crosslinker CL 210, inhibitor DVS, and platinum catalyst 510, also procured from Evonik, were each individually added and subsequently mixed to initiate the curing of the polymer. Finally, a vacuum stirrer was employed to ensure the liquid was bubble-free.

The finished liquid MAE was spread on a PET foil of 0.2 mm thickness using a film applicator with an adjustable blade set to the desired thickness. To accelerate the curing process, the sheet is placed in an oven for 1 h at 80°C, followed by 24 h at 60°C.

*Laser micromachining of MAE films*

The cured MAE was structured by laser micromachining, where material was selectively removed through controlled micro-ablation to form periodic lamellar patterns. The lamella height corresponded to the MAE layer thickness, while the width and periodicity were freely defined by the programmed laser path.

Laser structuring was carried out using a 20 W pulsed fiber laser (G4 Pulsed Fiber Laser, SPI Lasers UK Ltd.) operating at 1064 nm. The beam was directed by a galvanometric scanner (Raylase SS-IIE10) equipped with a telecentric f-θ lens (f = 56 mm, Ronar Smith), providing a working area of 20 × 20 mm². The incident beam diameter before the lens was 7 mm (measured at 1/e² of peak intensity), with a beam quality factor of M² = 1.3, resulting in an effective focal spot diameter of approximately 14 μm. The structuring pattern was generated using SAMLight software (SCAPS GmbH), allowing precise definition of lamella geometry.

A key advantage of this method lies in its self-limiting ablation behavior. By employing a substrate with low absorption at 1064 nm, ablation automatically ceased once the substrate surface get exposed, producing lamellae with nearly vertical sidewalls. Importantly, the process does not remove the embedded carbonyl iron particles from the remaining elastomer, thereby preserving the composite's magnetic responsiveness.

*Video-microscopy and video-image analysis*

The sample was mounted on a 3D-printed plastic stage with adjustable height in the direction perpendicular to the axis of the quadrupolar magnet. The individual lamellas were aligned parallel to the magnet axis (Figure 1b). The magnet was connected to a rotating unit driven by a stepper motor (Japan Servo Co. KP42HM1-017, 0.9° per step), which was controlled by an EasyDriver stepper motor driver board interfaced with an Arduino-based electronics platform. The control program was developed using the Arduino IDE. Typical operating voltages for the motor ranged from 15 to 30 V, with driving currents between 0.3 and 1 A.

The sample was illuminated from the front side using a multi-LED light source (GSVitec) with a luminous flux of 12000 lm. This high-intensity illumination was essential for capturing clear images at high frame rates using an ultrafast camera (PHOTRON FASTCAM NOVA S12), operating at 12,800 frames per second. Video analysis was performed using **Tracker**, an open-source video analysis and modeling tool. Tracker enables automatic tracking of defined regions of interest, e.g., a selected object in motion, and provides quantitative data on position, velocity, and acceleration.

**Acknowledgements**

This research was funded by the Slovenian Research and Innovation Agency (ARIS, research programs No. P1-0192 and P2-0231), the European Union's Horizon Europe research and innovation programme under Marie-Składowska-Curie Actions Doctoral Network MAESTRI (grant agreement No. 101119614), Slovenian-German bilateral research project (ARIS-DAAD, BI-DE/25-27-003, Project-ID:5773025), and within the framework of the GREENTECH project, co-financed by the European Union – NextGenerationEU. The work of R.K. and M.S. in Regensburg was funded by the German Research Foundation (DFG, project No. 524921900). R.K. and M.S. are grateful to Evonik Operations GmbH, Specialty Additives, Geesthacht, Germany, for providing chemicals for the synthesis of PDMS and to BASF SE, Ludwigshafen am Rhein, Germany, for providing CIP. The work was funded by the European Union. Views and opinions expressed are, however, those of the authors only and do not necessarily reflect those of the European Union or the European Research Executive Agency. Neither the European Union nor the granting authority can be held responsible for them.

**Conflict of Interest**

The authors declare no conflict of interest.

**Data Availability Statement**

All the graphs reported in Figures 1-8 (in TIFF format) and video-recordings S1 - S4 are available in the Repository of the University of Ljubljana under the PID: 20.500.12556/RUL-171495 (see also [24]).